\documentclass[twocolumn,showpacs,preprintnumbers,amsmath,amssymb]{revtex4}
\usepackage{amsfonts}
\usepackage{amssymb}
\usepackage{psfig}
\newcommand{\beq}{\begin{equation}}
\newcommand{\eeq}{\end{equation}}
\newcommand{\beqn}{\begin{eqnarray}}
\newcommand{\eeqn}{\end{eqnarray}}
\newcommand{\bea}[1]{\beq\begin{array}{#1}}
\newcommand{\eea}{\end{array}\eeq}

\newcommand{\Tr}[1]{\;{1\over #1}\mathop{\rm Tr}}
\newcommand{\tr}{\mathop{\rm Tr}}

\newcommand{\ket}[1]{|\,#1\,\rangle}
\newcommand{\bra}[1]{\langle\,#1\,|}
\newcommand{\braket}[2]{\langle\,#1\,|\,#2\,\rangle}

\newcommand{\diff}{\partial}
\newcommand{\cA}{{\cal A}}
\newcommand{\cB}{{\cal B}}

\newcommand{\cM}{{\cal M}}

\newcommand{\anzaz}{ansatz~}

\newcommand{\HP}[1]{{\mathrm{HP}^#1}}
\newcommand{\hp}{\mathrm{HP}}

\newcommand{\mean}[1]{{\langle #1 \rangle}}
\newcommand{\gone}{  0.27(3)}
\newcommand{\gtwo}{  0.27(3)}
\newcommand{\gthree}{0.27(4)}
\newcommand{\gfour}{0.27(4)}
\newcommand{\gfive}{0.26(2)}
\newcommand{\gsix}{ 0.27(3)}
\newcommand{\gseven}{ 0.28(3)}

\begin{document}
\preprint{ITEP-LAT/2005-30}

\title{On the Continuum Limit of Topological Charge Density Distribution}

\author{P.Yu.~Boyko}
  \email{boyko@itep.ru}
\author{F.V.~Gubarev}
  \email{gubarev@itep.ru}

\affiliation{Institute of Theoretical and  Experimental Physics,
             B.~Cheremushkinskaya 25, Moscow, 117218, Russia}

\begin{abstract}
The bulk distribution of the topological charge density, constructed
via $\HP{1}$ $\sigma$-model embedding method, is investigated.
We argue that the specific pattern of leading power corrections to gluon condensate
hints on a particular UV divergent structure of $\HP{1}$ $\sigma$-model fields,
which in turn implies the linear divergence of the corresponding topological density in the continuum limit. 
We show that under testable assumptions the topological charge is to be distributed within three-dimensional
sign-coherent domains and conversely, the dimensionality of sign-coherent regions dictates the leading divergence
of the topological density.
Confronting the proposed scenario with lattice data we present evidence for indeed peculiar divergence of the embedded fields.
Then the UV behavior of the topological density is studied directly and is found to agree with our proposition.
Finally, we introduce parameter-free method to investigate the dimensionality of relevant topological
fluctuations and show that indeed topological charge sign-coherent regions are likely to be three-dimensional.
\end{abstract}


\maketitle
\section{Introduction}
\label{section:intro}
Topology investigations had always been conspicuous topic for the lattice community
and the recent advances indeed put it on the solid grounds. Both the topological charge and its density
are now computable on thermalized vacuum configurations and
the results already obtained in pure Yang-Mills theories indicate that the conventional instanton
based models are to be strongly modified. Here we basically mean the discovery of global
topological charge sign-coherent regions~\cite{Horvath:2005rv}
and, what is even more important for us, the lower dimensionality of these
regions~\cite{Horvath:2005rv,Aubin:2004mp,Ilgenfritz:2005hh,Horvath:2005cv,Gubarev:2005jm,Zakharov:2006vt},
which are now believed to be three-dimensional.
Note that the lower dimensionality of physically relevant vacuum fluctuations should not come
completely unexpected, it had been repeatedly discussed in the recent past
(see, e.g., Refs.~\cite{Kovalenko:2005zx,Bornyakov:2001hs,Gubarev:2002ek}).

It is important that qualitatively the same picture of relevant topological fluctuations
appeared recently within radically different approach introduced and developed
in Refs.~\cite{Gubarev:2005rs,Boyko:2005rb}. Without mentioning all the details and technicalities
involved we note only that the $\HP{1}$ embedding method is the nearest to the classical ADHM
investigation of $SU(2)$ gauge fields topology and essentially reconstructs the topology defining
map $S^4\to \HP{1} = S^4$, in terms of which both the topological charge and its density obtain
unambiguous and well defined meaning.  We are not in the position to review all the
results obtained in~\cite{Gubarev:2005rs,Boyko:2005rb}, however, it is important
that $\HP{1}$ embedding allows to get rid of leading perturbative divergences
in various observables. Moreover, it reproduces the topological aspects
of $SU(2)$ Yang-Mills theory, is fairly compatible with other topology investigation
approaches and allows to calculate with amazing accuracy the gluon condensate,
demonstrating firmly that its quadratic power correction does not vanish.
It is crucial that the only non-trivial power dependence of $\HP{1}$ projected
curvature upon UV cutoff is contained in the quadratic term, which
must be encoded in the local structure of the topology defining map $S^4\to \HP{1}$.
In section~\ref{section:gluon} we argue that the only possibility to reproduce the observed 
pattern of power corrections is to assume that the mapping $S^4\to \HP{1}$ is highly asymmetric
so that the corresponding Jacobian (which is essentially equivalent to the topological density)
diverges linearly with diminishing lattice spacing.
In turn this divergence implies rather peculiar geometry of the relevant topological excitations.
Namely, we show in section~\ref{section:dimension} that the topological charge sign-coherent
regions are to be three-dimensional domains embedded into original four-dimensional space.
Here the consistency check is provided by  the fluctuations of topological
charges associated with sign-coherent regions.

Attempt to confront the above scenario with the lattice data brings out a wealth of both
technical and theoretical issues, which are addressed  in section~\ref{section:lattice}.
In section~\ref{section:eigen} we present numerical measurements which justify
our conclusion about the local structure of the map $S^4 \to \HP{1}$. 
The direct measurement of the UV behavior of the topological density (sections~\ref{section:cut}, \ref{section:qq})
necessitates the invention of new calculation algorithm, which turns out to be fast and rather accurate.
It allows us to make statistically significant comparison of the theoretical considerations
with lattice data and confirms that the characteristic topological density is indeed 
divergent, but at most linearly, in the continuum limit. As far as the dimensionality
of relevant fluctuations is concerned, we investigate it in section~\ref{section:diffusion}
and develop unambiguous, in fact, method of its determination. Essentially it is
the specially crafted biased random walk model embedded 
into the ambient topological density environment\footnote{
During this paper preparation we learned that similar in spirit ideas were discussed in Ref.~\cite{Gerrit}.
}. We argue that the appearance of critical-like behavior of the model signifies the lower
dimensional long-range order present in the topological density background.
We show that critical regime does occur and indicates that indeed the density is concentrated in highly
extended submanifolds, the dimensionality of which is compatible with three.
Finally we investigate the fluctuations of the topological charges associated with
sign-coherent regions and argue that only three-dimensional domains with
linearly divergent topological density are consistent with theoretical and experimental
restrictions imposed on the structure of topological fluctuations.

\section{Scaling of Topological Density}
\label{section:theory}

\subsection{Gluon Condensate, Leading Power Correction and Divergence of Topological Density}
\label{section:gluon}

The essence of $\HP{1}$ $\sigma$-model embedding approach is the assignment of unique configuration
of $\HP{1}$ $\sigma$-model fields $\ket{q_x}$ to every given SU(2) gauge background $A_\mu$
(until section~\ref{section:lattice} we use continuum notations), where $\ket{q_x}$ is two-component,
normalized $\braket{q_x}{q_x} = 1$, quaternionic vector (see Refs.~\cite{Gubarev:2005rs,Boyko:2005rb}
for details). The relevant configuration $\ket{q_x}$ is defined by the requirement that it
provides the absolute minimum to the functional
\beq
\label{gluon:func}
F(A, q) ~=~ \int d^4x\, \tr \left( A_\mu ~+~ \bra{q}\diff_\mu\ket{q} \right)^2
\eeq
for given (fixed) gauge potentials $A_\mu$. Note that gauge covariance is maintained
exactly since the $\sigma$-model target space $\HP{1}$ is the set of equivalence
classes with respect to $\ket{q_x} \sim \ket{q_x} \, \upsilon_x$, $\upsilon_x \in SU(2)$.
The uniqueness of the minimum of (\ref{gluon:func}) and the factual absence of Gribov copies problem
was discussed in length in Refs.~\cite{Gubarev:2005rs,Boyko:2005rb} and here we take it for granted.
Therefore the embedded $\HP{1}$ $\sigma$-model fields $\ket{q_x}$ are unique (although non-local) functions
of the original potentials. The advantage of the construction is that the gauge fields topology
becomes explicit in terms of $\ket{q_x}$ variables.
Indeed, the gauge invariant projectors $\ket{q_x} \bra{q_x}$ provide the map of compactified physical
space $S^4$ into the target space $\HP{1}$, the degree of which
is equal to the topological charge of the original gauge background.
Furthermore, the local distinction of this map from trivial one is the uniquely
defined measure of the topological charge density. One could say that the sole purpose
of $\HP{1}$ embedding method is to locally reconstruct the topology defining map $S^4 \to \HP{1}$,
which allows to get essentially all the topological aspects of the original background.

Then it is natural to introduce the $\HP{1}$ projection
\beq
\label{gluon:projection}
A_\mu ~\to ~ A^\hp_\mu ~\equiv~ - \bra{q}\diff_\mu\ket{q}\,,
\eeq
which replaces $A_\mu$ with its best possible approximation by $\sigma$-model induced potentials.
The striking properties of the projected fields $A^\hp_\mu$ were investigated in details in Ref.~\cite{Boyko:2005rb}.
In particular, it was shown that $A^\hp_\mu$ exactly reproduce the most of non-perturbative
aspects of the original $SU(2)$ configurations, while containing no sign whatsoever of usual perturbative
divergences. To the contrary, the kernel of the map (\ref{gluon:projection})
was shown to correspond to pure perturbation theory with identically trivial topology and
vanishing string tension. Without mentioning all the aspects and properties of $\HP{1}$ projection
(\ref{gluon:projection}) let us note that it allows to calculate with inaccessible so far accuracy
the gluon condensate and its leading power correction. Indeed, the spacing dependence of $\HP{1}$
projected gauge curvature $F^\hp_{\mu\nu}$ was found to be astonishingly well described by
\beq
\label{gluon:main}
\langle \Tr{2} (F^\hp_{\mu\nu})^2 \rangle ~=~ \frac{4 \alpha_2}{a^2} ~+~
\frac{\pi^2}{6} \, \langle \frac{\alpha_s}{\pi} G^2 \rangle\,,
\eeq
\beqn
\label{gluon:numbers}
\alpha_2 & = & [61(3)\mathrm{~MeV}]^2\,, \\
\langle \frac{\alpha_s}{\pi} G^2 \rangle\ & = & 0.0271(10)\mathrm{~GeV}^4\,, \nonumber
\eeqn
where $a$ is the lattice spacing ($1/a$ servers as the UV cutoff). It is crucial that Eq.~(\ref{gluon:main})
does not contain any sign of usual perturbative contribution of order $O(a^{-4})$.
Note that the value of $\alpha_2$ coefficient turns out
to be unexpectedly small, nevertheless it fits nicely into the known bounds on the magnitude of the quadratic
correction term (see Ref.~\cite{Rakow:2005yn} for recent review). While the actual numbers quoted
in (\ref{gluon:numbers}) are not important for the present discussion, it is crucial that they both are definitely
non-zero and therefore the spacing dependence of $\langle \tr (F^\hp_{\mu\nu})^2 \rangle$ includes only 
$O(a^{-2})$ and $O(a^0)$ terms.

To analyze the consequences of Eq.~(\ref{gluon:main}) let us consider point $n \in S^4$,
in the  neighborhood of which the map $S^4 \to \HP{1}$ is non-degenerate.
In the vicinity of $n$ and its image $m\in \HP{1}$ we introduce local coordinates $x^\mu$ and $y^\mu$, $\mu = 0,...,3$
(stereographic projection from points $n$ and $m$ correspondingly) such that $y_\mu(x=0) = 0$.
Non-degeneracy means that $\mathrm{det}[\diff y/\diff x] \ne 0$ and hence the function $y(x)$ is invertible.
From the specific explicit form of $\HP{1}$ projected fields (\ref{gluon:projection}) we conclude that
\beq
\label{gluon:inst}
A^\hp_\mu(x) = -\bra{q_x}\frac{\diff}{\diff x^\mu}\ket{q_x} = J^\nu_\mu(x) \, A^{inst}_\nu(y(x))\,,
\eeq
where $J^\mu_\nu = \diff y^\mu / \diff x^\nu$ is the Jacobian matrix, $A^{inst}_\mu(y)$
is the potential of classical BPST instanton solution with unit radius
\beq
A^{inst}_\mu(y) = - \bra{q} \diff_\mu \ket{q} = \frac{ (\bar e_\mu e_\nu - \bar e_\nu e_\mu) \, y^\nu}{ 2\,(1+y^2)}
\eeq
and $e_\mu$ denotes quaternionic basis.
Note that Eq.~(\ref{gluon:inst}) is local and crucially depends upon the space-time varying matrix $J^\nu_\mu$.
Furthermore, Eq.~(\ref{gluon:inst}) relies heavily on the $\HP{1}$ projection (\ref{gluon:projection}) and
would not be valid for generic gauge potentials.
The corresponding $\HP{1}$ projected curvature is similarly expressible in terms of instantonic field-strength
\beq
\label{gluon:F}
F^\hp_{\mu\nu} ~=~ J^{\rho_1}_\mu\,J^{\rho_2}_\nu \, F^{inst}_{\rho_1 \rho_2}\,,
\eeq
and therefore at point $n\in S^4$ we have
\beq
\label{gluon:map1}
\Tr{2} (F^\hp_{\mu\nu})^2 ~ \propto ~ (\tr g)^2 ~-~ \tr g^2\,,
\eeq
where we have introduced the metric $g^{\mu\nu} = J^\mu_\lambda J^\nu_\lambda$ and skipped inessential numerical factor. 
Therefore the study of the projected curvature $\mean{\Tr{2} (F^\hp_{\mu\nu})^2}$ is equivalent
to the local investigation of the metric $g$ associated with the topology defining map $S^4\to \HP{1}$.
In term of strictly positive eigenvalues $\lambda_\mu$ of $g$,
Eqs.~(\ref{gluon:main}) and (\ref{gluon:map1}) imply
\beqn
\label{gluon:map}
& \sum\limits_{\mu < \nu} \mean{\lambda_\mu \lambda_\nu} ~ = ~ A ~+~ B \cdot a^{-2}\,, & \\
& A \sim \Lambda^4_{QCD}\,, \quad B \sim \Lambda^2_{QCD}\,, & \nonumber
\eeqn
where we have generically indicated the IR physical scale involved in Eq.~(\ref{gluon:main}) by $\Lambda_{QCD}$
and explicitly kept all powers of lattice spacing.
Evidently, Eq.~(\ref{gluon:map}) imposes rather stringent restrictions on the distribution
of eigenvalues $\lambda_\mu$ and requires them to depend highly non-trivially upon the UV cutoff.
However, Eq.~(\ref{gluon:map}) is not sufficient to analyze this dependence in details.
The relation, which provides an additional input and which is verified numerically
in section~\ref{section:eigen}, reads
\beqn
\label{gluon:map2}
& \sum\limits_{\mu} \mean{\lambda_\mu} ~ = ~ \alpha ~+~ \beta \cdot a^{-2}\,, & \\
& \alpha \sim \Lambda^2_{QCD}\,,\quad \beta \sim 1\,. & \nonumber
\eeqn
Taken at face value it indicates that generically all the eigenvalues are quadratically divergent
in the limit $a\to 0$. However, it turns out that the simplest \anzaz
\beqn
\label{gluon:eigen}
& \mean{\lambda_0} = \beta \cdot a^{-2}\,, \quad \mean{\lambda_i} = \alpha_i\,,  \quad i = 1,2,3\,, & \\
& \alpha_i \sim \Lambda^2_{QCD} \,, \quad \sum_i \alpha_i = \alpha\,, & \nonumber
\eeqn
is capable not only to reproduce the observed pattern of power corrections (\ref{gluon:map}), (\ref{gluon:map2}),
but also passes stringent consistency check, which we describe next.
Note that without loss of generality the divergent behavior was ascribed to the first eigenvalue.
However, it is clear that  any particular enumeration of $\lambda_\mu$ has no invariant meaning.
What is actually meant in Eq.~(\ref{gluon:eigen}) is that only one eigenvalue is quadratically divergent,
but it is not possible to assign a particular number to it before averaging.
In order to convince the reader that (\ref{gluon:eigen}) is compatible with both 
(\ref{gluon:map}) and (\ref{gluon:map2}) we note that since $\lambda_0$ and $\lambda_i$
depend upon completely different scales it is legitimate to write
$\sum \mean{\lambda_\mu \lambda_\nu} = \mean{\lambda_0} \sum \mean{\lambda_i} + \sum \mean{\lambda_i \lambda_j}$.
Then Eq.~(\ref{gluon:map}) becomes
\beq
\alpha \, \beta \cdot a^{-2} + \sum\limits_{i < j} \mean{\lambda_i \lambda_j} = A + B \cdot a^{-2}
\eeq
and under quite natural (in view of (\ref{gluon:eigen})) assumption
$\sum \mean{\lambda_i \lambda_j} \sim \Lambda^4_{QCD}$ it leads
to indeed stringent relation between various coefficients
\beq
\label{gluon:check}
\alpha \cdot \beta ~=~ B\,.
\eeq
It is evident that this equality is highly non-trivial and is not guaranteed {\it a priori}.
If confirmed by lattice measurements it would imply the validity of the \anzaz (\ref{gluon:eigen})
thus providing the mean to check its self-consistency. As is discussed in section~\ref{section:eigen},
Eq.~(\ref{gluon:check}) is strongly supported by the measurements and is fulfilled with rather amazing accuracy.
Moreover, in that section we also consider along the same lines the triple correlator
$\sum_{\mu < \nu < \rho} \mean{\lambda_\mu \lambda_\nu \lambda_\rho}$, for which the analogous to (\ref{gluon:map})
relation holds and which provides the additional consistency check similar to (\ref{gluon:check}).
Note, however, that in the latter case the numerical uncertainties are larger and for this
reason we do not consider the triple correlator here.

To summarize, we found that the \anzaz (\ref{gluon:eigen}) reproduces precisely the observed
pattern of leading power divergences (\ref{gluon:main}), (\ref{gluon:map}), (\ref{gluon:map2})
and is fairly compatible with measured local characteristics of the topology defining map $S^4 \to \HP{1}$.
Since even the relation (\ref{gluon:check}) is strongly supported by the data,
we are confident that Eq.~(\ref{gluon:eigen}) reflects correctly the leading UV dependence
of the map $S^4 \to \HP{1}$, which therefore turns out to be highly asymmetric on average.
It is worth to note that the standard picture implies $\langle\lambda_\mu\rangle \propto \Lambda^2_{QCD}$
meaning that unit topological charge is gathered at large (of order $\Lambda^{-1}_{QCD}$) distances.
The singular behavior of one eigenvalue
$\langle\lambda_0\rangle \propto 1/a^2$ signifies immediately that the topological susceptibility
$\chi = \mean{Q^2}/V$ is saturated on submanifolds with characteristic four-volume of order
$a\cdot\Lambda^{-3}_{QCD}$. In other words the topological density is to be concentrated mostly
in three-dimensional domains embedded into Euclidean four-dimensional space
(we return to this problem in section~\ref{section:dimension}).

The above results have rather dramatic consequences for the topological density $q_x$,
to illustrate which let us consider $\mean{q^2}$ defined as $\mean{ q^2} = \lim_{|x|\to 0} \langle q_0 q_x\rangle$.
Note that within the usual approaches this definition is, in fact, ambiguous since the correlation function
$\mean{q_0 q_x}$ is perturbatively dominated at small distances $\langle q_0 q_x\rangle \sim -1/|x|^8$.
Hence the perturbative ambiguities make the value of $\langle  q^2 \rangle$ undefined
and the standard lore is to fix it by the requirement $\chi = \int \langle q_0 q_x\rangle$.
However, it is crucial that the $\HP{1}$ embedding approach to the gauge fields topology is factually
exempt from perturbation theory. The best illustration is provided by Eq.~(\ref{gluon:main}) and
in connection with topological density correlation function we discuss this in section~\ref{section:qq}.
Therefore for $\HP{1}$-based topological density, which in the continuum limit
is given by $q \propto \tr F^\hp \tilde F^\hp$, the estimate $\mean{q_0 q_x} \sim -1/|x|^8$
is not valid and UV behavior of $\mean{q^2}$ should be considered anew.
Note also that the possible UV divergence of $\mean{q^2}$ could not be subtracted as usual
and is not equivalent to conventional contact terms.
From now on and throughout the paper the topological density $q_x$ is
always understood via $\HP{1}$ embedding approach.

In fact, the leading UV behavior of $\HP{1}$-based topological density could be established
from Eq.~(\ref{gluon:F}). Indeed, from the well known properties of instanton solution it follows that 
\beq
\label{gluon:q}
q \propto \tr F^\hp \tilde F^\hp  \propto \mathrm{det}\, J =  \pm \mathrm{det}^{1/2}\, g\,.
\eeq
Here it is convenient to introduce the notion of characteristic topological density $\bar q$, which could
be defined rigorously as $\bar q^2 = \mean{q^2}$. However, below we'll sometimes use this term and the same
symbol $\bar q$ to denote just the typical scale of the topological density fluctuations.
The justification is that as far as the dependence on UV and IR scales is concerned these definitions
essentially coincide. Therefore, from Eq.~(\ref{gluon:q}) we conclude that 
\beq
\label{density:spacing}
\bar q^2 \equiv \langle  q^2 \rangle \propto \langle \mathrm{det}\, g \rangle =
\langle\prod\limits_\mu \lambda_\mu \rangle \sim \frac{\Lambda^6_{QCD}}{a^2}\,,
\eeq
where Eq.~(\ref{gluon:eigen}) had been used.
The conclusion is that even the non-perturbatively defined topological density is divergent in the continuum limit
reflecting directly the highly asymmetric local structure of the topology defining map $S^4 \to \HP{1}$.
Note however that this divergence is incomparable with usual perturbative
one $\langle  q^2 \rangle \sim 1/a^8$, which, in fact, is even non-integrable.
Of course, it is understood that the very definition of the topological density is 
arbitrary to large extent (full derivative could always be added).
However, the $\HP{1}$ embedding method is specified completely with no free parameters involved.
Moreover, the corresponding topological density is definitely exempt from perturbative ambiguities
so that the divergence (\ref{density:spacing}) could not be equivalent to the contact term and should be dealt with accurately.
In particular, we argue in the next section that Eq.~(\ref{density:spacing}) 
implies rather peculiar geometry of vacuum topological fluctuations.

\subsection{Dimensionality of Topological Fluctuations}
\label{section:dimension}
The problem to be addressed in this section is the geometrical properties of vacuum
topological fluctuations and in particular their dimensionality.  First, let us note that
any particular distribution of topological density in finite volume $V$
could be divided unambiguously into the regions $V_+$ ($q_x>0$) and $V_-$ ($q_x<0$)
of sign-coherent topological charge so that the total charge is given by
\beq
\label{dimension:decompose}
Q = \int q = \int_{V_+}\!\! q  ~+ \int_{V_-} \!\! q \equiv Q_+ - Q_-\,,
\eeq
where the relative sign of $Q_+$, $Q_-$ was made explicit in the last equality ($Q_- > 0$)
and no specific properties of the regions $V_+$, $V_-$ were supposed
($V_\pm$ could be disconnected, in particular).
Note that in this section we do assume that the charges $Q_\pm$ could be estimated as
$Q_\pm = \bar q \, V_\pm$. For mean squared topological charge the decomposition (\ref{dimension:decompose}) implies
\beq
\label{dimension:Q2}
\mean{Q^2} ~\propto~ \mean{Q^2_+} - \mean{Q_+ Q_-}\,,
\eeq
where we have generically assumed that $\mean{Q^2_-}=\mean{Q^2_+}$.
Since the topological susceptibility $\chi$ is {\it a priori} postulated to be finite
in the continuum limit, we have
\beq
\label{dimension:Q2-2}
\frac{\mean{Q^2_+} - \mean{Q_+ Q_-}}{V} =
\bar q^2 \frac{\mean{V^2_+} - \mean{V_+ V_-}}{V} \sim  \Lambda^4_{QCD}\,.
\eeq
Keeping in mind the ultraviolet divergence of $\bar q$, Eq.~(\ref{density:spacing}),
we conclude that the independent fluctuations of the
volumes $V_\pm$ are strictly prohibited. Instead the magnitudes of $V_+$ and $V_-$
are to be tuned up to the order $O(a^2)$
\beq
\label{dimension:fine-tuning}
\mean{ (V_+ - V_-)^2 } ~\sim~  a^2 \, \Lambda^{-2}_{QCD} \cdot V \,,
\eeq
since otherwise the topological susceptibility would diverge in the continuum limit.
It is important that, contrary to the case of zero-point fluctuations,
for non-perturbatively defined topological density
there are no arguments which would guarantee the exact generic cancellation of divergent
terms in the integral $Q = \int q$.
In fact, the fine tuning assumption (\ref{dimension:fine-tuning}) is not new,
similar in spirit observations were made already in the recent past
(see, e.g., Refs.~\cite{Gubarev:2005jm,Bornyakov:2001hs,Gubarev:2002ek,Kovalenko:2005rz}).
However, we would like to reformulate the problem so that explicit powers of lattice spacing do not appear.
Indeed, Eq.~(\ref{dimension:fine-tuning}) implicitly assumes that the volumes $V_\pm$  are four-dimensional
and is satisfied identically  if the topological density is distributed in three-dimensional
domains $V^{(3)}_\pm = V_\pm/a$, which are allowed to fluctuate on the scale of $\Lambda_{QCD}$
\beq
\label{dimension:three}
\mean{ (V^{(3)}_+ - V^{(3)}_-)^2 } ~\sim~  V / \Lambda^2_{QCD}\,.
\eeq
Note that (\ref{dimension:three}) is not the real solution, but rather the reformulation,
of the fine tuning problem. Indeed, although the explicit spacing dependence is gone,
the three-dimensional structure of topological fluctuations in D=4 YM theory is equivalent, in fact,
to a sort of fine tuning.

In order to make the presentation more coherent, let us return to Eq.~(\ref{dimension:Q2}).
We could easily bypass Eq.~(\ref{dimension:fine-tuning}) assuming that the charges $Q_\pm$
fluctuate independently
\beq
\label{dimension:1}
\mean{Q_+ Q_-} = \mean{Q_+} \mean{Q_-} = \mean{Q_+}^2\,.
\eeq
Then Eq.~(\ref{dimension:Q2-2}) translates into
\beq
\frac{\mean{Q^2_+} - \mean{Q_+}^2}{V} \propto \frac{\mean{Q_+}}{V} =
\bar q \,\,\frac{\mean{V_+}}{V} \sim \Lambda^4_{QCD}\,,
\eeq
where the validity of central limit estimate 
\beq
\label{dimension:2}
\mean{Q^2_+} - \mean{Q_+}^2 \propto \mean{Q_+}
\eeq
was supposed. Then the relation between ultraviolet behavior
of characteristic topological density and the dimensionality of the corresponding fluctuations
could be  given as follows. Assume that $\bar q$ is of order $a^{-\alpha}$ and that the dimensionality
of the topological fluctuations is $D$ so that $V_+ = a^{4-D} \cdot V^{(D)}_+$, where $V^{(D)}_+$
is spacing independent. Then
\beq
\chi \sim a^{4 - \alpha - D} \cdot \frac{\mean{V^{(D)}_+}}{V} \sim \Lambda^4_{QCD}
\eeq
and the relation between UV behavior of the characteristic topological density and the
dimensionality of the relevant topological fluctuations follows
\beq
\label{dimension:relation}
\bar q ~\sim~ a^{-\alpha}\,, \qquad \mathrm{dim}[V_\pm] ~=~ 4 - \alpha\,.
\eeq
However, it is clear that the argumentation relies heavily on the assumption that the fluctuations
of the topological charges $Q_\pm$ obey Eq.~(\ref{dimension:1}) (as well as Eq.~(\ref{dimension:2}),
which, however, seems to be less restrictive).
{\it A priori} Eq.~(\ref{dimension:1}) is by no means evident and being confronted with experimental lattice data
provides the most stringent test of the above scenario. Various experimental aspects of the problem
are addressed in the next section. 
Here we note only that Eqs.~(\ref{density:spacing}), (\ref{dimension:1})
imply the three-dimensional structure of vacuum topological fluctuations. Evidently, the reversed
argumentation could also be given, namely, the dimensionality of sign-coherent topological charge
fluctuations determines the leading ultraviolet behavior of characteristic topological density
provided that Eq.~(\ref{dimension:1}) is valid.
We stress that the essence of the above presentation is the factual absence of leading perturbative divergences
in $\HP{1}$ projected fields and in the corresponding topological density.
One could convince oneself that in the case of perturbatively dominated topological density, $\bar q \sim a^{-4}$,
the assumption (\ref{dimension:fine-tuning}) (with $a^2$ replaced by $a^4$ on the r.h.s.) holds true,
while Eq.~(\ref{dimension:1}) is violated and reads instead $\mean{Q_+ Q_-} = \mean{Q^2_\pm}$.
The conclusion is that  Eqs.~(\ref{density:spacing}), (\ref{dimension:1}), (\ref{dimension:2})
are indeed crucial to validate the lower dimensionality of the topological charge fluctuations.

\section{Confronting with Lattice Data}
\label{section:lattice}
In this section we describe in details the results of our numerical investigations
of the scenario outlined above.
In section~\ref{section:eigen} we study the lattice spacing dependence of the eigenvalues
of the metric associated with the topology defining map $S^4 \to \HP{1}$ and discuss in details
the results announced in section~\ref{section:gluon}.
In section~\ref{section:cut} the topological density at various scales is considered;
we show that even the simplest approach indeed qualitatively confirms the divergence
of characteristic topological density in the continuum limit.
Section~\ref{section:qq} is devoted to the investigation of $\mean{q_0 q_x}$ correlation
function from which we deduce the scaling law of the characteristic topological density.
Then in section~\ref{section:diffusion} we propose a method, which includes essentially
no free parameters and allows to directly establish the dimensionality of topological fluctuations.
Finally in section~\ref{section:check} the topological charges associated with 
sign-coherent domains are shown to fluctuate indeed independently thus providing
the self-consistency check of the above scenario.

\begin{table}[t]
\centerline{\begin{tabular}{|c|c|c|c|c|c|c|} \hline
$\beta$ & $a$,fm & $L_t$ & $L_s$ & $V^{phys}, \mathrm{fm}^4$ & $N^{conf}$ & $N^{conf}_q$\\ \hline
 2.4000 & 0.1193(9)  &  16 & 16 & 13.3(4) & 198 & 70 \\
 2.4273 & 0.1083(15) &  16 & 12 &  3.8(2) & 250 & 80 \\
 2.4500 & 0.0996(22) &  14 & 14 &  3.8(2) & 200 & 80 \\
 2.4750 & 0.0913(6)  &  16 & 16 &  4.6(1) & 380 & 75 \\
 2.5000 & 0.0854(4)  &  18 & 16 & 3.92(7) & 200 & 75 \\
 2.5550 & 0.0704(9)  &  20 & 20 &  3.9(2) &  80 & 80 \\
 2.6000 & 0.0601(3)  &  28 & 28 &  8.0(2) &  65 & 60  \\ \hline
\end{tabular}}
\caption{Simulation parameters.}
\label{tab:params}
\end{table}

The numerical measurements were performed on 7 sets (Table~\ref{tab:params})
of statistically independent SU(2) gauge configurations generated with standard Wilson action. 
The most of configurations listed in Table~\ref{tab:params} are the same as were used in
Refs.~\cite{Gubarev:2005rs,Boyko:2005rb} (except for the set at $\beta = 2.555$).
The last column in Table~\ref{tab:params} represents the number of configurations on which
we calculated the bulk topological charge density. Note that the number of analyzed
configurations at each spacing is indeed rather large, which is due to the new algorithm
used to evaluate the topological density (the algorithm is described in Appendix).
The lattice spacing values quoted in Table~\ref{tab:params} were partially taken from
Refs.~\cite{spacings} and fixed by the physical value of SU(2) string tension
$\sqrt{\sigma} = 440~\mathrm{MeV}$. Note that for $\beta=2.4273$ and $\beta=2.555$
the lattice spacings and corresponding rather conservative error estimates were obtained
via interpolation in between the points quoted in~\cite{spacings}.

\subsection{Local structure of the map $S^4\to \HP{1}$}
\label{section:eigen}
The local structure of $S^4\to \HP{1}$ mapping is characterized by the corresponding
Jacobian $J^\mu_\nu$ (see (\ref{gluon:inst}), (\ref{gluon:F}), (\ref{gluon:q})), however
in this section we concentrate on the induced metric $g^{\mu\nu} = J^\mu_\lambda J^\nu_\lambda$
and its strictly positive eigenvalues $\lambda_\mu$, $\mu=0,...,3$.
As we noted already, any particular enumeration of $\lambda_\mu$ has no invariant meaning
so that the meaningful observables associated with $g^{\mu\nu}$ can not depend upon the ordering.
Let us first describe the actual numerical procedure
used to extract the spectrum $\{\lambda_\mu\}$ at any particular lattice point $x$, at which
we know the unit five-dimensional vector $n^A_x \in \HP{1}$, $A=0,...,4$ as well as the analogous
quantities at the neighboring sites $n^A_{x+\nu} \in \HP{1}$.
In accord with what had been said in section~\ref{section:gluon} we introduce stereographically
projected coordinates $y^\mu_x$, $y^\mu_{x+\nu}$ such that $y^\mu_x = 0$ and then consider the
discretized approximation to the Jacobian
\beq
J^\mu_\nu(x) ~=~ a^{-1} \cdot (y^\mu_{x+\nu} ~-~ y^\mu_x) ~=~ a^{-1} \cdot y^\mu_{x+\nu}\,,
\eeq
from which the metric and its spectrum $\{\lambda_\mu\}$ are obtained straightforwardly.
It goes without saying that we confronted the quantity
$[\sum_{\mu < \nu } \lambda_\mu \lambda_\nu](x) $ with $[\Tr{2} (F^\hp_{\mu\nu})^2](x)$
on each configuration and found that Eq.~(\ref{gluon:map1}) is satisfied almost identically
at every lattice site.

\begin{figure}[t]
\centerline{\psfig{file=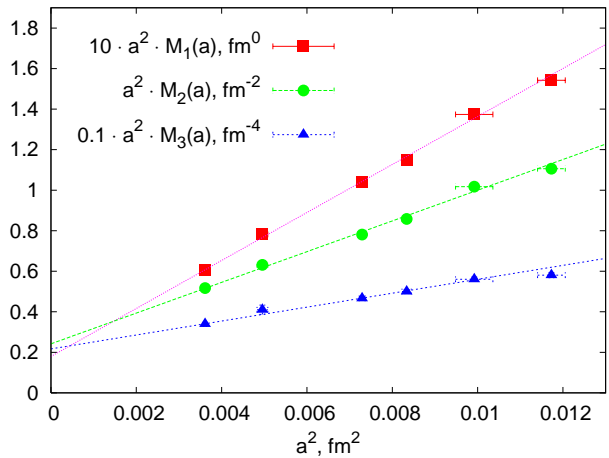,width=0.5\textwidth,silent=,angle=-90}}
\caption{Scaling of the correlators $\cM_n$, Eqs.~(\ref{eigen:m1}), (\ref{eigen:m2}), (\ref{eigen:m3}),
with diminishing lattice spacing. Lines represent the best fits according to Eq.~(\ref{eigen:moments}).}
\label{fig:moments}
\end{figure}

The lattice spacing dependence was measured for three ordering insensitive
observables associated with $g^{\mu\nu}$
\beq
\label{eigen:m1}
\cM_1(a) = \sum\limits_{\mu} \mean{\lambda_\mu}\,,
\eeq
\beq
\label{eigen:m2}
\cM_2(a) = \sum\limits_{\mu < \nu } \mean{\lambda_\mu \lambda_\nu}\,,
\eeq
\beq
\label{eigen:m3}
\cM_3(a) = \sum\limits_{\mu < \nu < \rho} \mean{\lambda_\mu \lambda_\nu \lambda_\rho}\,.
\eeq
The results of our measurements are presented on Fig.~\ref{fig:moments}
and indicate strongly that UV behavior of all these quantities is well described by
\beqn
\label{eigen:moments}
& \cM_n ~=~ \alpha_n  ~+~ \beta_n \cdot a^{-2}\,, \quad n = 1,2,3 \,, & \\
& \alpha_n \sim \Lambda^{2n}_{QCD}\,, \quad \beta_n \sim \Lambda^{2(n-1)}_{QCD}\,. & \nonumber
\eeqn
It is remarkable that the observed pattern of power corrections is universal and for each correlator
(\ref{eigen:m1}), (\ref{eigen:m2}), (\ref{eigen:m3}) includes only terms of order $O(a^0)$ and $O(a^{-2})$.

What is even more important here is that the numerical data (\ref{eigen:moments}) is in perfect qualitative
agreement with the assumption (\ref{gluon:eigen}). Moreover, the actual numerical values of the coefficients
$\alpha_n$, $\beta_n$ provide the self-consistency check of the \anzaz (\ref{gluon:eigen}). Indeed,
considering Eq.~(\ref{eigen:moments}) for $n=1$ and $n=2$ we obtain
\beqn
\label{eigen:check1}
\alpha_1 \cdot \beta_1  & = &  \beta_2\,, \\
\sum\limits_{ i < j } \mean{\lambda_i \lambda_j }  & = & \alpha_2\,, \nonumber
\eeqn
while the combination of $n=2$ and $n=3$ cases leads to
\beq
\label{eigen:check2}
\beta_3 ~=~ \beta_1 \cdot \sum\limits_{ i < j } \mean{\lambda_i \lambda_j } ~=~ \beta_1 \cdot \alpha_2\,.
\eeq
Evidently, Eqs.~(\ref{eigen:check1}), (\ref{eigen:check2}) are highly non-trivial and their numerical validity
would definitely signify the correctness of the \anzaz (\ref{gluon:eigen}).
As far as the numerical values of the relevant coefficients are concerned,
the outcome of the best fits according to Eq.~(\ref{eigen:moments}) is
\beq
\begin{array}{rcrcrcr}
a^2 \cdot \alpha_1 & = & 11.8(3)\,, &  &           \beta_1 & = & 0.018(2)\,, \\
a^4 \cdot \alpha_2 & = &  87(3)\,,   &  & a^2 \cdot \beta_2 & = &  0.24(2)\,, \\
                   &   &            &  & a^4 \cdot \beta_3 & = & 2.1(2)\,.
\end{array}
\eeq
One can see that these numbers are in the perfect agreement with Eq.~(\ref{eigen:check1})
\beq
\frac{\alpha_1 \cdot \beta_1}{ \beta_2 } ~=~ 0.9(1)\,,
\eeq
and are compatible with Eq.~(\ref{eigen:check2})
\beq
\frac{\beta_1 \cdot \alpha_2}{\beta_3} ~=~ 0.8(2)\,,
\eeq
although the deviation from unity and numerical uncertainty is larger in the latter case.

The conclusion is that numerical data supports strongly the assumption (\ref{gluon:eigen})
so that indeed the topology defining map $S^4 \to \HP{1}$ is likely to be highly asymmetric.
One of the relevant eigenvalues is divergent in the continuum limit
and seems to depend only on  the ultraviolet cutoff. 
On the other hand, the remaining eigenvalues are sensitive solely
to the infrared scale and show no divergences near the continuum limit. 

Finally, we note that it is tempting to consider along the same lines the fourth order
correlator $\cM_4  = \mean{\prod \lambda_\mu}$ and then relate via Eq.~(\ref{density:spacing})
its spacing dependence with leading UV behavior of the topological density.
In turns out that numerically $\cM_4(a)$ is indeed compatible with (\ref{eigen:moments}),
however, we refrain to rely on (\ref{density:spacing}) at finite lattice spacing
The point is that Eq.~(\ref{density:spacing}) is certainly valid in the limit $a\to 0$,
provided  that $\cM_4$ stays constant in physical units.
However, due to the suspected power divergence the corrections to Eq.~(\ref{density:spacing})
are difficult to estimate.

\subsection{Topological Fluctuations at Various Scales}
\label{section:cut}
As was repeatedly stressed in Refs.~\cite{Gubarev:2005rs,Boyko:2005rb}
any discussion of the topological charge density within the lattice settings inevitably
introduces a particular cutoff $\Lambda_q$ on the magnitude of the density
so that $q(x)$ is equated to zero if $|q(x)| < \Lambda_q$.
Indeed, the most straightforward argument here is that in the numerical simulations the
density is always known with finite accuracy. Thus the numerical precision
provides the finest possible cutoff which in physical units evidently scales like
$\Lambda_q \propto a^{-4}$. Moreover, the introduction of (often implicit) finite $\Lambda_q$
is inherent to all studies of the gauge fields topology. For instance, the chiral fermions based topological
density, which is given by the sum of Dirac eigenmodes $\psi_\lambda$ contributions,
is usually either restricted to lowest modes, $\lambda < \Lambda \propto \Lambda_{QCD}$, or is considered
for all modes available on the lattice, $\lambda \lesssim 1/a$.
Therefore the actual problem is not the presence of the cut $\Lambda_q$, it is
introduced always. The physically meaningful question is the spacing dependence
$\Lambda_q = \Lambda_q(a)$ and the above
examples illustrate two extreme cases $\Lambda_q \propto a^{-4}$ and $\Lambda_q \propto \Lambda^4_{QCD}$.

It is crucial that the scaling law $\Lambda_q(a)$ could be taken at will and we're going to exploit
this freedom to study the spacing dependence of the topological density. Indeed, if the characteristic
topological density $\bar q$ stays constant in physical units then the volume density
of points at which $|q(x)| > \Lambda_q$,
\beq
\label{cut:rho}
\rho(\Lambda_q) = \frac{1}{V} \sum\limits_x \left\{ \begin{array}{cc} 1\,, & |q(x)| > \Lambda_q \\
0\,, & \mbox{otherwise} \end{array} \right. \,,
\eeq
should also be lattice spacing independent for $\Lambda_q \propto \Lambda^4_{QCD}$.
Note that the lattice units had been used in (\ref{cut:rho}) and that
$\rho(\Lambda_q)$ is dimensionless, positive, bounded $\rho(\Lambda_q) \le 1$ quantity,
unrelated to the volumes $V_\pm$ discussed in section~\ref{section:dimension}.
It is clear that the divergence $\bar q \sim a^{-\alpha}$ would result in the divergent behavior
$\rho(\Lambda_q) \sim a^{-\alpha}$ of the volume density\footnote{
Note that in the numerical simulations the singularity should not be expected,
the divergence is only seen for $\rho \ll 1$.
} for physical cut $\Lambda_q \propto \Lambda^4_{QCD}$, while $\rho(\Lambda_q)$ is to be
almost spacing independent for similarly divergent cut
$\Lambda_q \propto \Lambda^4_{QCD} \cdot (a \, \Lambda_{QCD})^{-\alpha}$.
Therefore the most straightforward way to analyze the spacing dependence of the characteristic
topological density is to tune the scaling law $\Lambda_q (a)$ until the volume density
$\rho(\Lambda_q)$ becomes constant at various lattice resolutions.

\begin{figure}[t]
\centerline{\psfig{file=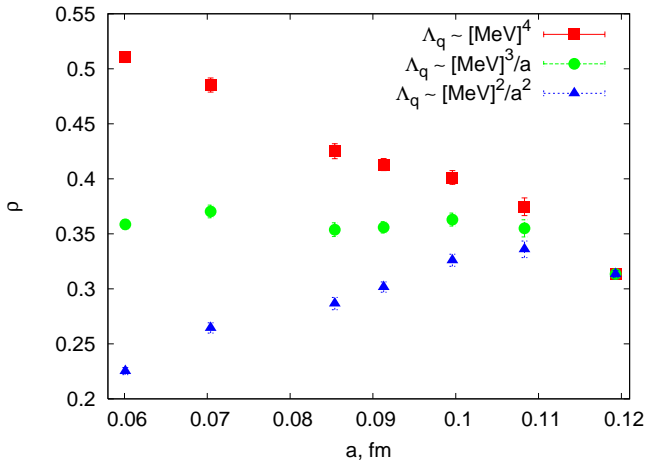,width=0.5\textwidth,silent=,angle=-90}}
\caption{Volume density of points with $|q_x| > \Lambda_q$ as a function of lattice spacing
at different $\Lambda_q$ scaling laws (\ref{cut:scaling-law}).}
\label{fig:volume-density}
\end{figure}

Unfortunately, this approach does not allow to investigate the dependence $\bar q(a)$ precisely.
Indeed, on the lattice we could only probe a finite set of scaling laws $\Lambda_q(a)$, moreover
the corresponding estimates of $\rho(\Lambda_q)$ are always biased. However, it is crucial
that the correct qualitative picture could easily be obtained this way. We performed
the measurements of the volume density $\rho(\Lambda_q)$ at various lattice spacings
using three different scaling laws
\beq
\label{cut:scaling-law}
\Lambda^{(n)}_q ~ \propto ~ \Lambda^4_{QCD} \cdot (a\,\Lambda_{QCD})^{-n}\,,\qquad n = 0, 1, 2\,,
\eeq
where the numerical coefficients were chosen in such a way that
$\Lambda^{(n)}_q = [200 \mathrm{~MeV}]^4$, $n=0,1,2$ at $a = 0.1193\mathrm{~fm}$
(this particular choice is motivated below).
The results of our measurements are presented on Fig.~\ref{fig:volume-density} from which
it is clear that the volume density of points with $|q_x| > \Lambda^{(0)}_q = [200\mathrm{~MeV}]^4$
is rapidly rising with diminishing lattice spacing.
Contrary to that the linearly divergent cut on the topological density, $\Lambda^{(1)}_q \sim 1/a$
results in the almost spacing independent volume density $\rho(\Lambda^{(1)}_q) \approx 0.35$.
On the other hand, once the quadratically divergent cut $\Lambda^{(2)}_q \sim 1/a^2$ is imposed
the quantity $\rho(\Lambda^{(2)}_q)$ diminishes almost linearly with vanishing lattice spacing
and in the limit $a\to 0$ becomes compatible with zero.
The conclusion is that the characteristic magnitude of the topological density
is indeed singular in the continuum limit, the leading divergence is compatible with linear one
\beq
\label{cut:divergence}
\bar q ~\propto ~ \Lambda^3_{QCD}\,/\,a
\eeq
and is in accord with theoretical expectations (\ref{density:spacing}).

Let us discuss now the dimensionality of topological charge sign-coherent regions.
Qualitatively the structure of topological fluctuations at various cuts $\Lambda_q$ is rather
simple~\cite{Gubarev:2005rs,Boyko:2005rb}. For utterly small values of $\Lambda_q$
there are typically only two large (percolating) regions of sign-coherent topological density,
each of which occupies almost half of the lattice volume and carries rather large topological charge
$Q_\pm$, Eq.~(\ref{dimension:decompose}). With rising cutoff the volume density of percolating regions
diminishes, while the number of small sign-coherent lumps rapidly grows. Finally a sort of percolation
transition happens at which the percolating regions cease to exist
and become indistinguishable from the small lumps.
After that point the volume distribution of sign-coherent regions
become universal ($\Lambda_q$ independent)  and is described by rather remarkable power law.
However, the physics changes drastically at the lumps percolation transition.
Namely, the string tension, associated with $\HP{1}$ projected fields and
which accounts for the full $SU(2)$ string tension in the continuum limit, vanishes.
Already from this observation we expect that the most physically important topological
fluctuations are represented by the largest (at given cutoff $\Lambda_q$) lumps in topological density
and it is natural to focus exclusively on their dimensionality.
Note that the actual values of the cuts (\ref{cut:scaling-law}) were taken to be
always below the lumps percolation transition in the whole range of lattice spacings considered.

However, at any fixed cutoff $\Lambda_q$ the structure of the lumps is very complicated and
their dimension is, in fact, not a well defined concept. Indeed, the notion of dimensionality makes sense
only as a scaling relation since at any fixed lattice spacing the lumps occupy some finite
fraction of the volume. Actually the situation is much worse since the very definition of the lumps
require introduction of the cutoff $\Lambda_q$, the spacing dependence of which is not fixed.
Moreover, admitting the lower dimensionality of sign-coherent regions, their
volume fraction is not obliged to be finite in the continuum limit, hence
even the spacing dependence of lumps localization degree (which might be expressed
in terms of inverse participation ratio or similar quantities) would not reveal their dimensionality.

It is clear that the crucial obstacle in lumps dimensionality definition is the necessity to impose
the cutoff on the topological density. The concept of the dimensionality of topological fluctuations
would become unambiguous provided that we could get rid of explicit $\Lambda_q$ and hence reject the language
of the lumps. This program is implemented in section~\ref{section:diffusion}.
However before going into details let us study the divergence (\ref{cut:divergence})
more quantitatively and consider the topological density correlation function.

\subsection{$\langle q_0 q_x \rangle$ Correlation Function}
\label{section:qq}
It is was discussed in brief in section~\ref{section:gluon} that considering the magnitude of the characteristic
topological density defined by $\bar q^2 \equiv \lim_{x \to 0} \mean{q_0 q_x}$ one has to prove that $\bar q$
indeed makes sense and is not equivalent to the contact term inherent to the perturbation theory.
It was stated without proof that this is the case for $\HP{1}$-based topological density.
In this section we present the corresponding data
and investigate the correlation function $\mean{q_0 q_x}$. Note that 
the correlator $\mean{q_0 q_x}$ is known to be negative at any non-vanishing
distance~\cite{Seiler:2001je} provided that the definition of the topological density is local
(see also Refs.~\cite{Ilgenfritz:2005hh,Horvath:2005cv,Horvath:2005rv} for discussions).
However, the requirement of locality is {\it a priori} violated in $\HP{1}$ embedding
approach so that $\mean{q_0 q_x}$, $|x| \ne 0$ is not obliged
to be negative. We could only hope that the intrinsic non-locality
is not so violent and extends up to some distance $R_0$ fixed in physical units.
Note that this expectation is not completely groundless.
Indeed, many non-perturbative observables, defined via $\HP{1}$ projection and studied
in Refs.~\cite{Gubarev:2005rs,Boyko:2005rb}, do not reveal any pathology
and reproduce, in fact, the corresponding results in the full theory.
Actually the degree of non-locality of $\HP{1}$ method could be estimated from the behavior
of heavy quark potential measured at $a = 0.0601\mathrm{~fm}$ in~\cite{Boyko:2005rb}
and it turns out to be of order $R_0 \lesssim 0.2\mathrm{~fm}$. On the other hand, the investigation
of $\mean{q_0 q_x}$ correlation function allows to find $R_0$ rather precisely and
check its scaling properties.

\begin{figure}[t]
\centerline{\psfig{file=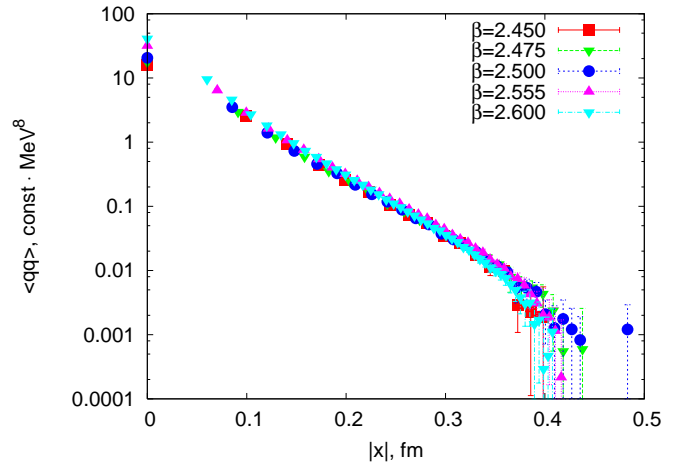,width=0.5\textwidth,silent=,angle=-90}}
\caption{Positive core of topological density correlation function $\mean{q_0 q_x}$
at distances $|x| < R_0 \approx 0.4 \mathrm{~fm}$.}
\label{fig:qq-core}
\end{figure}

\begin{figure}[t]
\centerline{\psfig{file=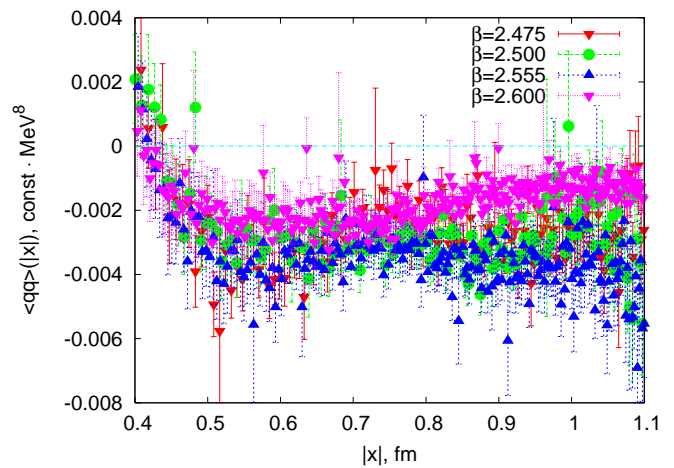,width=0.5\textwidth,silent=,angle=-90}}
\caption{Negative part of the correlation function $\mean{q_0 q_x}$ at distances
$|x| \ge R_0 = 0.4\mathrm{~fm}$ at various lattice spacings.}
\label{fig:qq-negative}
\end{figure}

Generically we expect that the correlator $\mean{q_0 q_x}$ is to be positive up to
the distance $R_0$ and then should
become negative provided that $R_0$ is finite. Remarkably enough these expectations are precisely
confirmed by the measurements. The positive core of $\mean{q_0 q_x}$ correlation function
at small $|x|$ is presented in physical units on Fig.~\ref{fig:qq-core}. It is apparent that
the points at various spacings are well described by the exponential $|x|$-dependence
\beq
\label{qq:exp}
\mean{q_0 q_x} ~\propto ~ e^{-|x|/R_{qq}}\,, \qquad |x| \lesssim R_0\,.
\eeq
Note that due to the logarithmic scale used on this plot the data sets are terminating
at the same physical distance, for larger $|x|$ the correlation function becomes negative.
Therefore the degree of non-locality inherent to $\HP{1}$ embedding method is
given roughly by
\beq
\label{qq:nonlocal}
R_0 ~\lesssim~ 0.4\mathrm{~fm}
\eeq
and is indeed constant in physical units as is evident from Fig.~\ref{fig:qq-core}.
Moreover, the almost perfect scaling of various data sets indicates once again
that $\HP{1}$ projected fields do not contain any trace of the perturbation theory.
Indeed, the mixture with perturbative contributions would lead to notable $\sim - 1/|x|^8$
terms and would result in rather abrupt deviation from the exponential behavior.
On the other hand, at large distances, $|x| \gtrsim R_0$, the correlation function $\mean{q_0 q_x}$
indeed becomes negative as is illustrated on Fig.~\ref{fig:qq-negative}. It is important that the
negative part of $\mean{q_0 q_x}$ correlator does not show any singularity in the limit $a\to 0$,
in particular, it has nothing to do with usual perturbative dependence.

\begin{figure}[t]
\centerline{\psfig{file=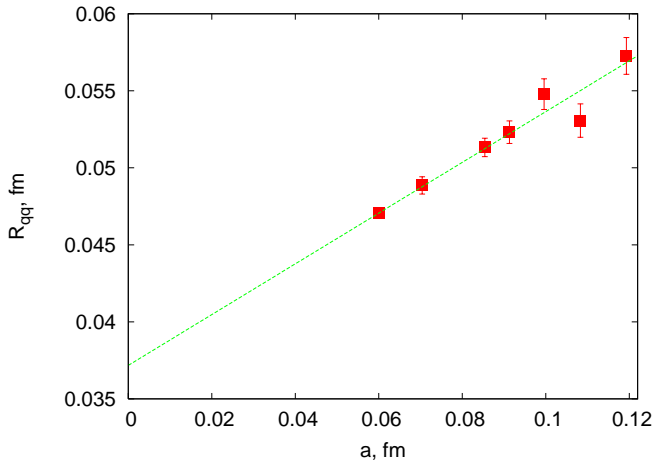,width=0.5\textwidth,silent=,angle=-90}}
\caption{Correlation length $R_{qq}$, Eq.~(\ref{qq:exp}), as a function of lattice spacing.
Line represents the best linear fit.}
\label{fig:qq-length}
\end{figure}

Let us consider the scaling properties of the correlation length $R_{qq}$, Eq.~(\ref{qq:exp}).
As might be expected already from Fig.~\ref{fig:qq-core}, $R_{qq}$ decreases with diminishing lattice spacing.
In more details the dependence $R_{qq}(a)$ is presented on Fig.~\ref{fig:qq-length}, from
which it is apparent that the correlation length is likely to be linear function of $a$ with
rather small continuum value
\beq
\label{qq:length}
R_{qq} ~=~ 0.037(1)\mathrm{~fm}\,.
\eeq

\begin{figure}[t]
\centerline{\psfig{file=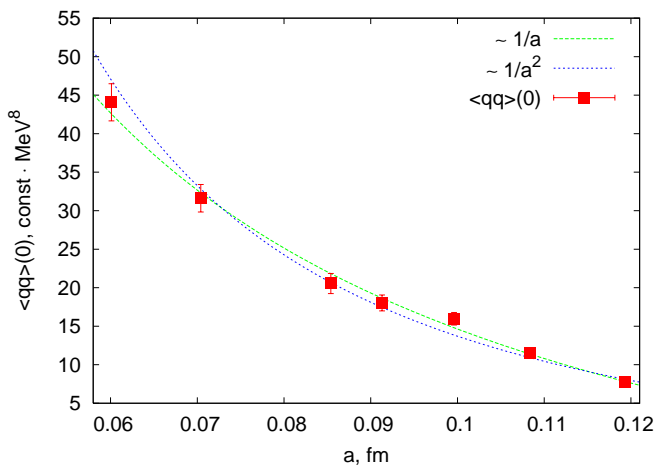,width=0.5\textwidth,silent=,angle=-90}}
\caption{The mean squared topological density $\mean{q^2}$ as a function of lattice spacing.
Curves represent the best power law fits, Eq.~(\ref{qq:fits}).}
\label{fig:qq-zero}
\end{figure}

As is apparent from the above presentation, the short distance behavior of $\mean{q_0 q_x}$ correlation function 
is definitely exempt from the perturbative uncertainties and in the limit $|x|\to 0$ the characteristic topological
density
\beq
\bar q^2 = \mean{q^2} = \lim_{|x|\to 0} \mean{q_0 q_x}
\eeq
has nothing to do with contact terms inherent to the usual approaches. 
Therefore let us consider the dependence of $\bar q$ upon the ultraviolet cutoff.
The results of our measurements are presented on Fig.~\ref{fig:qq-zero} from which
it is clear that $\HP{1}$-based characteristic topological density still diverges in the continuum limit. 
However, this divergence has nothing in common with perturbatively
expected $O(a^{-8})$ behavior, in fact, it is much weaker and is compatible only with linear or
quadratic dependence
\beq
\label{qq:fits}
\mean{q^2} ~=~ b_n ~+~ c_n \cdot a^{-n}\,, \qquad n = 1,2\,.
\eeq
Unfortunately, our data points do not distinguish these power laws and are adequately
described by either linear ($\chi^2_{n=1} = 0.9$) or quadratic ($\chi^2_{n=2} = 1.5$) one.
Nevertheless, the ultraviolet divergence of characteristic topological density $\bar q \sim a^{-\alpha}$
within the $\HP{1}$ embedding method could be considered as firmly established.
Moreover, we are confident that the corresponding power
exponent is close to unity, $\alpha \lesssim 1$, and is in accord with theoretical expectations
(\ref{density:spacing}) and the estimate (\ref{cut:divergence}) obtained earlier.

\subsection{Dimensionality of Topological Charge Fluctuations}
\label{section:diffusion}

\subsubsection{Biased Random Walk Model and the Choice of Parameters}
\label{section:model}

In this section we introduce the model, which allows to investigate directly the dimensionality
of the relevant topological fluctuations. 
Although the resulting method is not entirely rigorous, its ambiguity reduces to
only one free parameter, the choice of which we thoroughly discuss.
Generically the idea is to consider some dynamical system the evolution of which is sensitive
to the dimensionality of the ambient space. Then if we embed somehow this system into
the topological density background, its evolution will reveal the effective number of
available dimensions, which is to be naturally associated with the dimensionality
of the relevant topological fluctuations. The simplest dynamical system of this sort 
could be constructed on the top of usual diffusion equation, which in turn is equivalent
to the model of random walks. Then the dependence upon the external environment could be introduced
by making the hopping probabilities to be the local functions of the ambient space characteristics;
the models of this sort are known as biased random walks.
Correspondingly, the effective dimensionality $D$ as is seen by biased random walkers 
is referred to as diffusion dimension. The purpose of this section is to precisely formulate
and investigate this approach.

We start from the observation that the introduction of the cutoff $\Lambda_q$ on the topological density
(see section~\ref{section:cut}) was aimed solely to separate the inevitably
present noise (utterly small values of $q_x$) from the relevant
fluctuations, which are associated with relatively large values of $q_x$.
Although the choice of particular $\Lambda_q$ indeed makes the notion of 'small' and 'large'
well defined, the geometry of the resulting lumps in topological density strongly depends upon the cut.
It is apparent that the weak point here is sharpness of the dimension four cut,
it would be much advantageous to remove the small topological density regions softly,
making the lumps geometry much more robust with respect to the parameters involved.
It turns out that slightly modified diffusion model is indeed suitable to achieve this.
Namely, we propose to modify the diffusion equation by allowing
the random walk to hop towards the regions of higher topological density with larger probability.
In the language of the diffusion equation, which describes the propagation of heat, this amounts to the
introduction of space-time dependent diffusion coefficient (thermal conductivity), which vanishes in the regions of
small topological density, and hence heat is allowed to spread only within the domains of large $|q_x|$.
Then the decay rate of the initial heat pulse, which is the same as the return probability for corresponding
biased random walk, essentially reflects the number of available dimensions within the
topological fluctuations and hence is to be identified with their dimensionality.

In fact, this general idea fixes almost uniquely the random walk model which we would like to investigate.
It is convenient to start directly from the microscopic rules of the biased random walk, which
require that the probability $p_{x, x+\mu}$ to hop from point $x$ to the neighboring site $x+\mu$
is monotonically rising scale free function of the topological density magnitude at $x+\mu$
\beq
\label{diffusion:def}
p_{x, x+\mu} = 
\frac{|q_{x+\mu}|^\gamma}{\sum_\mu (|q_{x+\mu}|^\gamma + |q_{x-\mu}|^\gamma)}\,,
\eeq
where the power exponent $\gamma > 0$ could not be fixed {\it a priori} and remains free parameter of the model.
Note that in this section we exclusively consider the local magnitude $|q_x|$ of the topological density;
to lighten the notations the corresponding modulus sign will be omitted.
From Eq.~(\ref{diffusion:def}) it is straightforward to obtain the continuum diffusion-like
equation, which determines the probability $P(t,x)$ to reach the point $x$ during the proper time
interval $t$ provided that at $t=0$ walker starts at $x=0$ (see, e.g., Ref.~\cite{Itzykson})
\beqn
\label{diffusion:eq-x}
& \diff_t \Phi ~=~ \frac{1}{8} \, q^{-2\gamma} \, \diff_x \left[\,q^{2\gamma}\, \diff_x \Phi\right]\,, & \\
& \Phi(t,x) \equiv P(t,x)/q^{2\gamma}(x)\,, \nonumber
\eeqn
where the initial condition is $P(0,x) = \delta(x)$.
However, Eq.~(\ref{diffusion:eq-x}) is not yet the usual diffusion equation, in particular,
the decay rate of the initial perturbation (return probability in the random walk language) is not 
guaranteed to be $P(t,0) \propto t^{-D/2}$. To get the correct interpretation of (\ref{diffusion:eq-x})
we introduce new coordinates $\zeta(x)$ according to
\beq
\label{diffusion:change}
\diff \zeta^\mu / \diff x^\nu ~=~ q^{2\gamma} \cdot \delta^\mu_\nu\,.
\eeq
In the particular case of one dimensional problem (\ref{diffusion:change}) allows explicit solution
$\zeta(x) = \int^x q^{2\gamma}$, where the lower integration limit is taken at arbitrary fixed point.
It is important that
$\zeta$ is a single valued function of $x$ almost everywhere, moreover, its range is determined by
the magnitude of the topological density. Indeed, the regions of utterly small $|q_x|$ 
are squeezed to almost one point by the map (\ref{diffusion:change}) regardless how large these regions
were in $x$ space. It is crucial that the term 'small' above obtains unambiguous 
and physically correct meaning of relative smallness since
only the relative variation of the topological density does matter.
Indeed, Eq.~(\ref{diffusion:eq-x}) is evidently scale invariant under $q \to \lambda q$ and is
equivalent to usual diffusion equation for everywhere constant $q_x$.
In terms of new coordinates  Eq.~(\ref{diffusion:eq-x}) takes the standard form
\beq
\label{diffusion:eq-zeta}
\diff_t \Phi ~=~ \frac{1}{8} \, \diff_\zeta \left[\,q^{4\gamma}\, \diff_\zeta \Phi\right]\,.
\eeq
We conclude therefore that the diffusion process 
(\ref{diffusion:def}), (\ref{diffusion:eq-x}), (\ref{diffusion:eq-zeta}) takes place in the regions
of relatively large topological density and hence should reflect properly the dimensionality of underlying topological
background. Moreover, the decay rate of the initial perturbation is given by
\beq
\label{diffusion:rate}
\Phi(t, 0 ) ~ \propto ~ t^{-D(\gamma)/2}\,,
\eeq
where $D(\gamma)$ is the diffusion dimension of the topological fluctuations
and we have explicitly indicated that the dimension defined this way depends non-trivially upon
yet not fixed parameter $\gamma$ to be discussed next.

The non-triviality of the dependence $D(\gamma)$ is evident since in the limit $\gamma \to 0$
the model (\ref{diffusion:def}), (\ref{diffusion:eq-x}) reduces  to standard unbiased random walk with
\beq
\label{diffusion:trivial}
D(\gamma \to 0) ~=~ 4\,,
\eeq
while at $\gamma \to \infty$ the microscopic probability (\ref{diffusion:def})
allows the hopping only towards largest neighboring $|q_x|$.
Hence the walker is trapped eventually at the local maxima of topological density distribution and
\beq
\label{diffusion:trivial2}
D(\gamma \to \infty) ~=~ 0\,,
\eeq
regardless of the background.
Already from this observation it is apparent that without an additional physical input
the model (\ref{diffusion:def}), (\ref{diffusion:eq-x}) would be essentially useless
since at various $\gamma$ it reflects, in fact, different properties of the underlying background.
For instance, the limiting behavior (\ref{diffusion:trivial}) implies that the topological density
distribution is such that $q_x \equiv 0$ only on measure zero set, while (\ref{diffusion:trivial2}) 
is valid generically, provided $q_x$ is not identically constant.
Therefore the actual problem is not the arbitrariness of $\gamma$ parameter, but rather yet not
posed physical question we're trying to investigate.

In order to gain a physical insight it is crucial to retain the qualitative
picture of vacuum topological excitations outlined in section~\ref{section:cut}.
It was argued that physically most important fluctuations are associated with percolating 
sign-coherent regions, the ultimate qualitative properties of which
are the significant internal topological density and extremely large linear extent.
It goes without saying that we are interested precisely in the geometry of these sign-coherent domains. 
However, the percolating lumps are not equivalent to just the regions of largest topological density
as is revealed by the percolation transition eventually appearing with rising $\Lambda_q$ cutoff.
Even at large $\Lambda_q$ one finds
individual 'hot spots' of small but non-vanishing volume which indeed possess the largest topological density.
Consider now the decay rate of the initial heat pulse in the model (\ref{diffusion:eq-x}) at large but finite
$\gamma$, so that the effective thermal conductivity is non-zero only within the topological density hot spots.
Evidently, in this regime the equilibration time $t_{eq}$ is finite and is dictated by the typical size of the
hot spots being much smaller than the squared lattice size $L$
\beq
\label{diffusion:gamma_large}
t_{eq} \ll L^2 \quad \mbox{for} \quad \gamma \gg \gamma^*
\eeq
(the definition of $\gamma^*$ will become clear in the moment).
Note that the distinct feature of this regime is that the double logarithmic plot of $\Phi(t,0)$ 
significantly bends upwards at $t_{eq}$ and therefore we expect generically that
\beq
\label{diffusion:log}
\gamma > \gamma^* \,: \quad
\frac{\diff^2 \ln \Phi(t,0)}{ \diff (\ln t)^2} > 0\,,
\eeq
reflecting, in particular, the drop in the effective thermal conductivity outside the hot spots.
Evidently, at these $\gamma$ values we are not probing the percolating lumps geometry,
the random walks are confined within the regions of largest  topological density.

Let us now gradually diminish the $\gamma$ parameter.  The positive jump in the logarithmic derivative
$\diff \ln \Phi / \diff \ln t$, the location of which we still denote by $t_{eq}$, would become smaller
respecting the diminishing difference of thermal conductivities inside and outside the hot spots.
Note that neither of the corresponding diffusion dimensions at $t \gtrless t_{eq}$ could be identified
with the dimensionality of sign-coherent regions since the random walks are still too sensitive to the
local irregularities of topological density even within the coherent domains.
It might happen that at the particular value $\gamma = \gamma^*$ the double logarithmic plot of $\Phi(t,0)$ 
degenerates into the straight line so that $t_{eq} \approx L^2$ and
\beq
\label{diffusion:log2}
\gamma = \gamma^* \,: \quad \frac{\diff^2 \ln \Phi(t,0)}{ \diff (\ln t)^2} = 0\,,\quad t \lesssim L^2\,.
\eeq
Note that at this point the only relevant dimensional parameter is the lattice size.
Therefore the model becomes essentially scale free and its dynamics in the vicinity
of $\gamma^*$ is reminiscent to the disorder driven conductor-insulator transition of condensed matter systems.
It is crucial that at the critical point $\gamma = \gamma^*$ the heat starts to propagate with constant rate
over largest available distances, but since the thermal conductivity is significant only within the lumps,
the heat transfer goes through the percolating sing-coherent regions.
In turn the condition (\ref{diffusion:log2}) implies that local inhomogeneity of the topological density
within the percolating lumps is inessential.
Therefore it seems that only at the critical point (\ref{diffusion:log2})
we indeed  could obtain  a consistent reflection of the relevant
topological fluctuations in the biased random walks model.

The existence of at least one critical value, $\gamma^* = 0$, follows from (\ref{diffusion:trivial}),
but it is trivial one and surely exists for arbitrary background $q_x$.
However, it is crucial that the above qualitative picture of vacuum topological fluctuations
hints on the existence of non-trivial critical coupling $\gamma^* >0$, which could arise entirely
due to the low-dimensional long-range order present in the topological density distribution.
For suppose that the topological charge sign-coherent regions are indeed lower dimensional objects
possessing relatively large uniform topological density and extending through all the volume. 
Then at the particular $\gamma = \gamma^*$  the effective thermal conductivity would be non-zero only
within the percolating regions, the lower dimensionality of which forbids the appearance
of an additional scale in the heat propagation problem. Thus at this point the random walks model
indeed contains no dimensional parameter apart from the lattice size and Eq.~(\ref{diffusion:rate})
is fulfilled, while the corresponding diffusion dimension is to be identified naturally
with the dimension of sign-coherent regions.
Note that the lower dimensionality is crucial here, for regions of finite thickness
the critical point (\ref{diffusion:log2}) does not exists.
In fact, the analogous but reversed argumentation could also be given, namely, the appearance
of non-trivial critical point (\ref{diffusion:log2}) signifies the presence of lower dimensional
long-range order in the topological density background. 

Note that the restriction to sign-coherent regions is automatic in our approach. Indeed, as far as the
model (\ref{diffusion:eq-x}) is concerned, it was considered in the continuum limit assuming differentiability
of $q_x$. Hence the domains $q_x \gtrless 0$ are separated by regions with vanishing thermal conductivity.
On the lattice the situation is more involved, but since the $\HP{1}$-based topological density
is definitely exempt from perturbative contributions we could be almost confident that the same
argumentation applies.

To summarize, the proposed biased random walk model seems to be efficient
to investigate  the structure of percolating topological density regions  
only at the critical points, at which its dynamics becomes effectively scale free.
We argued that the qualitative picture of the relevant topological fluctuations
obtained earlier suggests the existence of non-trivial critical point, at which
the diffusion dimension is to be identified naturally with the percolating regions dimensionality.

\subsubsection{Diffusion Dimension of Percolating Lumps}
\label{section:diffdim}

Prior to presenting the results of our measurements let us discuss the lattice specific
features of the above biased random walk model, which complicate the numerical evaluation
of the percolating lumps dimensionality.
For any particular $\gamma$ value the solution of Eq.~(\ref{diffusion:eq-x})
could be constructed straightforwardly by implementing the random walk
process, the rules of which are completely specified by (\ref{diffusion:def}).
The only subtlety here is the choice of the random walk starting point.
Indeed, to improve the statistics it is desirable to consider
\beqn
\label{diffusion:init}
\Phi(t, 0) = \frac{\int d\zeta_0 \, \Phi(0, \zeta_0 ; t, \zeta_0)}{\int d\zeta_0} = \\
= \int dx_0 \, \Phi(0, x_0; t, x_0 ) \, \frac{q^{8\gamma}(x_0)}{\int q^{8\gamma}} \,, \nonumber
\eeqn
where $\Phi(0, x_0;t, x)$ is the probability to reach the point $x$ during the proper time $t$
starting at $x_0$. Eq.~(\ref{diffusion:init}) means that the random walk starting point
in $x$ space is to be taken with probability  $\propto q^{8\gamma}$.

\begin{figure}[t]
\centerline{\psfig{file=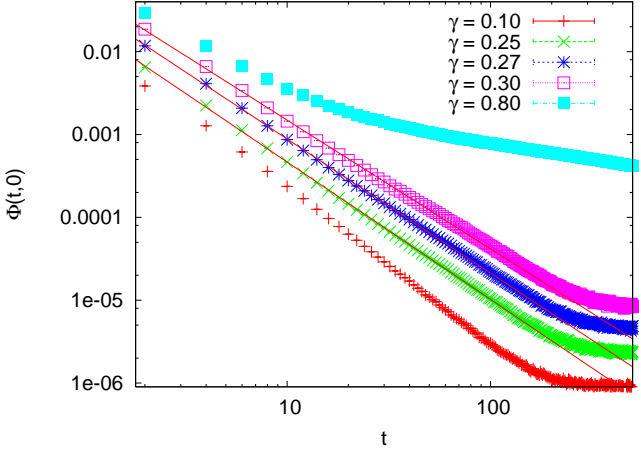,width=0.5\textwidth,silent=,angle=-90}}
\caption{Return probability of biased random walk $\Phi(t,0)$, Eqs.~(\ref{diffusion:def}, \ref{diffusion:eq-x}),
at different $\gamma$ values for $\beta = 2.500$ data set. Solid lines are the best fits of
$\gamma = \{ 0.25, 0.27, 0.30 \}$ data points to Eq.~(\ref{diffusion:rate}), fitting range is $t < 100]$.}
\label{fig:deviation}
\end{figure}
\begin{figure}[t]
\centerline{\psfig{file=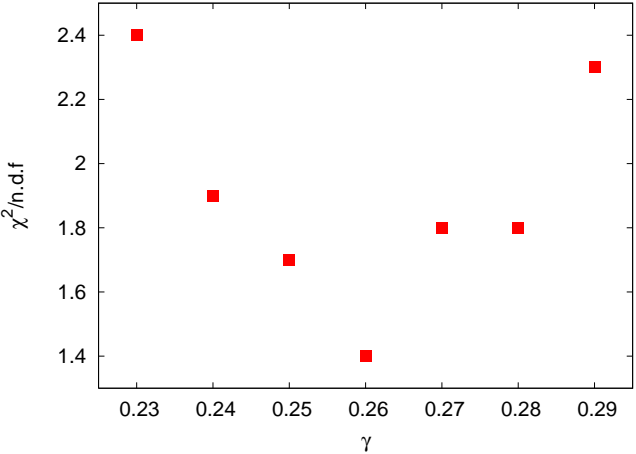,width=0.5\textwidth,silent=,angle=-90}}
\caption{Quality of the power law fits of the return probability $\Phi(t,0)$ to Eq.~(\ref{diffusion:rate})
in the range $t < 100$ at various $\gamma$ values. Measurements were performed on $\beta = 2.500$ data set.}
\label{fig:chi}
\end{figure}

Therefore the crucial problem is how to find the relevant $\gamma^*$ value numerically
and estimate the accuracy of its determination.
For any fixed $\gamma$ we measured $\Phi(t, 0)$ by first taking some random starting point,
chosen with $\propto q^{8\gamma}$ probability,
and considering in accord with (\ref{diffusion:def}) the walk of total length $8 \cdot 10^3$. 
Then the quantity $\Phi(t,x) = P(t,x)/q^2(x)$ was constructed for each random walk and averaged
with respect to $\approx V/2$ different starting points per configuration ($V$ is the lattice volume).
We checked that the statistics is large enough so that our results do not change
if more random walks are considered.
Fig.~\ref{fig:deviation} represents the double logarithmic plot of the quantity $\Phi(t,0)$
obtained on our $\beta=2.500$ set at three close $\gamma$ values, $\gamma = \{0.25, 0.27, 0.30\}$,
as well as the same quantity measured at $\gamma = 0.1$ and $\gamma = 0.8$.
Note that for readability reasons different graphs are slightly shifted with respect to each other.
As far as the data points at $\gamma = 0.1$ and $\gamma = 0.8$ are concerned they definitely correspond 
to the regimes $\gamma \ll \gamma^*$ and $\gamma \gg \gamma^*$ respectively. Indeed, the first
graph significantly bends downwards at $t \approx 10 \div 20$ in apparent disagreement with both
Eq.~(\ref{diffusion:log}) and Eq.~(\ref{diffusion:log2}).
However, it is demonstrated below that negative second logarithmic derivative of $\Phi(t,0)$
is a generic feature of the diffusion model in random environment.
Note that at $\gamma = 0.1$ the system indeed equilibrates eventually at $t_{eq}$ of order few hundred,
which is similar to squared size of the lattice used in this calculation.
On the other hand, at $\gamma = 0.8$
the system starts to equilibrate at $t_{eq} \approx 20 \ll L^2$  and the corresponding graph 
has positive second derivative in accord with Eq.~(\ref{diffusion:log}).

\begin{figure}[t]
\centerline{\psfig{file=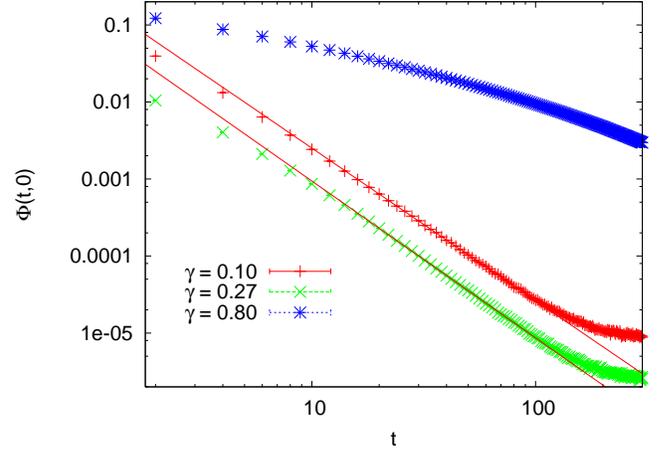,width=0.5\textwidth,silent=,angle=-90}}
\caption{Return probability $\Phi(t,0)$, Eqs.~(\ref{diffusion:def}, \ref{diffusion:eq-x}),
calculated on $\beta=2.500$ data set with random permutation of topological density records
on each configuration.}
\label{fig:random}
\end{figure}
\begin{table}[t]
\centerline{\begin{tabular}{|c|c|c|c|c|c|c|c|} \hline
$\beta$    &  2.4000 & 2.4273 & 2.4500  & 2.4750 & 2.5000     & 2.5550 & 2.6000 \\ \hline
$\gamma^*$ &  \gone  & \gtwo  & \gthree & \gfour & \gfive     & \gsix  & \gseven \\ \hline
\end{tabular}}
\caption{Estimated $\gamma^*$ values for various data sets from Table~\ref{tab:params}.}
\label{tab:gamma}
\end{table}

The inspection of the intermediate $\gamma$ values reveals that $\gamma = 0.25$ data points essentially
lie on one single line, while the data set at $\gamma = 0.30$ seems to deviate upwards from linear dependence.
Note that the consideration of $\gamma=0.27$ points are rather inconclusive, apparently the data just begins
to bend aside the pure power law.
Therefore the relevant $\gamma^*$ value seems to be located around $\gamma^* \approx 0.25$.
We could estimate $\gamma^*$ more rigorously  by considering the quality of power law (\ref{diffusion:rate})
fits at different $\gamma$.
We fitted our data to Eq.~(\ref{diffusion:rate}) in the range $t < 100$, the resulting $\chi^2/\mathrm{n.d.f.}$ values
are presented on Fig.~\ref{fig:chi}. It is apparent from this figure that fits favor the critical value
$\gamma^* = \gfive$, where rather conservative errors coming from Fig.~\ref{fig:chi} are quoted.

In order to convince the reader that graphs on Fig.~\ref{fig:deviation} are indeed directly related
to the underlying geometry of topological fluctuations, let us consider the same model
(\ref{diffusion:def}), (\ref{diffusion:eq-x}) with the same parameters,
but in the genuinely random external environment.
Such a background, which is the 'nearest' to one on Fig.~\ref{fig:deviation},
could be obtained by random permutation of the topological density records on each configuration.
We did this with our $\beta = 2.500$ data set and then performed the identical measurements of $\Phi(t,0)$
at $\gamma = \{0.10, 0.27, 0.80\}$.
The corresponding double logarithmic plots are presented on Fig.~\ref{fig:random} and differ violently
from that on Fig.~\ref{fig:deviation}. The most crucial observation is that the second derivative for all graphs
stays negative implying that the reasoning of section~\ref{section:model} is inapplicable for randomly
permuted topological density. Of course, this is not surprising since all the coherent structures
are gone in the present case. The only limit in which the second logarithmic derivative of $\Phi(t,0)$
vanishes is given either by Eq.~(\ref{diffusion:trivial}), $D=4$, or Eq.~(\ref{diffusion:trivial2}), $D=0$.
We conclude therefore that without random permutations the non-trivial critical point $\gamma^*$
arises most likely due to the presence of long-range order in the topological density.

We performed the measurements of the relevant $\gamma^*$  values for all our data sets
listed in Table~\ref{tab:params}.
It goes without saying that indeed in each case the biased random walk return probability
at $\gamma = \gamma^*$ strictly obeys the power law (\ref{diffusion:rate}).
The resulting estimations of corresponding $\gamma^*$ values are summarized in Table~\ref{tab:gamma}.
It is remarkable that the estimates of $\gamma^*$ appear to be spacing independent well within
the quoted numerical uncertainties. It is true that the presented errors might be overestimated,
however, we are confident that $\gamma^*(a)$ dependence could not be revealed by the above method.
Apparently this is due to the fact that $\gamma$ parameter is dimensionless so that the violent
power dependence on the lattice spacing is unlikely to appear.

Once the relevant $\gamma^*$ values are determined for every our data set it is straightforward
to estimate the dimensionality of sign-coherent percolating regions of the topological density.
By construction the corresponding fits to Eq.~(\ref{diffusion:rate}) are practically perfect for every
$\beta$ so that we do not present full details of the fitting procedure.
The point which should be discussed, however, is $D(\gamma^*)$ error estimation. 
In fact, the uncertainty coming from the fits to Eq.~(\ref{diffusion:rate}) is negligible compared to the ambiguity
in $\gamma^*$ values. Therefore the errors in $D(\gamma^*)$ were obtained by repeating the fits
to Eq.~(\ref{diffusion:rate}) at minimal and maximal $\gamma^*$ within the corresponding error bands.
Finally, we note that all fits were performed in the range $t < 100$.

\begin{figure}[t]
\centerline{\psfig{file=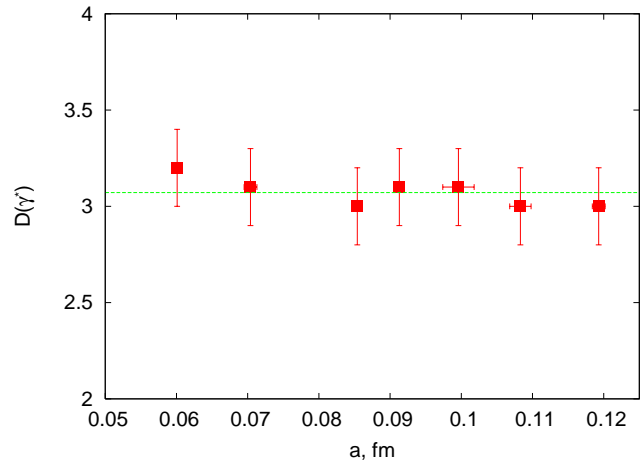,width=0.5\textwidth,silent=,angle=-90}}
\caption{Diffusion dimensions $D(\gamma^*)$, Eqs.~(\ref{diffusion:def}), (\ref{diffusion:eq-x}), (\ref{diffusion:rate}),
at various lattice spacings. Line represents the fit to the constant behavior.}
\label{fig:diffusion}
\end{figure}

The dimensionality of percolating sign-coherent regions of topological density
measured with the above procedure at various lattice spacings is summarized 
on Fig.~\ref{fig:diffusion}.
Remarkably enough the diffusion dimension $D(\gamma^*)$ seems to be  almost independent upon
the lattice resolution well within the uncertainties.
Its continuum value could be estimated from fit to constant behavior,
which indicates in turn that $D(\gamma^*)$ is definitely smaller than 4.
Instead the numerical data suggests strongly that the diffusion dimension of the relevant
topological fluctuations is
\beq
\label{diffusion:dim}
D ~ \equiv ~ D(\gamma^*) ~=~ 3.07(3)\,,
\eeq
which seems to be in full agreement with their proposed three dimensional structure.
Eq.~(\ref{diffusion:dim}) is in agreement with both theoretical expectations
and experimental data on the topological density. What is still to be considered is the internal consistency
of the scenario outlined in our paper, which is expressed by Eqs.~(\ref{dimension:1}),
(\ref{dimension:2}); this is the subject of the next section.
Note, however, that the diffusion dimension is not the only definition of the dimensionality.
It is still worth to confirm Eq.~(\ref{diffusion:dim}) with other methods.

\subsection{Consistency Check}
\label{section:check}

\begin{figure}[t]
\centerline{\psfig{file=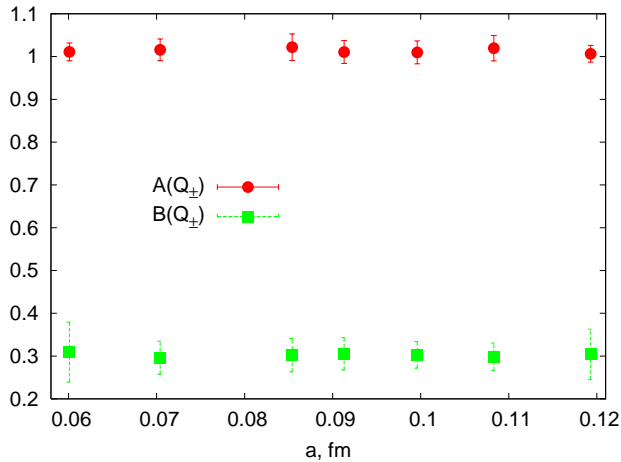,width=0.5\textwidth,silent=,angle=-90}}
\caption{Fluctuations of the topological charges $Q_\pm$ at various lattice spacings.
Circles: spacing (in)dependence of the ratio $\cA(Q_\pm)$, Eq.~(\ref{check:fig});
squares: the magnitude of $Q_\pm$ relative fluctuations characterized by $\cB(Q_\pm)$, Eq.~(\ref{check:fig}).}
\label{fig:check}
\end{figure}

As was discussed in section~\ref{section:dimension} the crucial equations which relate the divergence
of characteristic topological density $\bar q$ and the dimensionality $D$  of the relevant 
topological fluctuations are
\beq
\label{check:main1}
\mean{Q_+ Q_-} = \mean{Q_+} \mean{Q_-}\,,
\eeq
\beq
\label{check:main2}
\mean{Q^2_\pm} - \mean{Q_\pm}^2 \propto \mean{Q_\pm}\,,
\eeq
where the charges $Q_\pm$ are to be calculated without any cutoff $\Lambda_q$ imposed.
Moreover, the spacing independence of (\ref{check:main1}), (\ref{check:main2}) 
(if confirmed by the data) allows to put rather stringent restrictions
on both $\bar q$ and $D$.
As far as the lattice measurements are concerned, Fig.~\ref{fig:check} represents the ratios
\beqn
\label{check:fig}
\cA(Q_\pm) = \frac{\mean{Q_+ Q_-}}{\mean{Q_+}\mean{Q_-}} & ~ & \mbox{(circles)}\,, \\
\cB(Q_\pm) = \frac{\mean{Q^2_\pm} - \mean{Q_\pm}^2}{\mean{Q_\pm} } & ~ & \mbox{(squares)}\,, \nonumber
\eeqn
as a functions of lattice spacing, where to improve the statistics the generic equalities
$\mean{Q^2_+} = \mean{Q^2_-}$, $\mean{Q_+} = \mean{Q_-}$ were used in calculation of $\cB(Q_\pm)$.
As is evident from that figure Eq.~(\ref{check:main1}) is satisfied identically
in the whole range of considered spacings
while the proportionality coefficient entering Eq.~(\ref{check:main2})
does not depend upon the lattice resolution well within numerical errors.
Therefore the validity of Eq.~(\ref{dimension:relation}) is firmly established.
Let us summarize the emerging qualitative picture of vacuum topological fluctuations 
which arises from the numerical data restricted by (\ref{dimension:relation}).

It is apparent that the most confidential data is available for the characteristic topological density.
The theoretical arguments based on the existence of the quadratic correction to the gluon condensate as well as
the obtained numerical data suggest strongly that the topological density is divergent at most
linearly in the continuum limit
\beq
\label{check:q}
\bar q ~\sim ~ a^{-\alpha}\,, \qquad \alpha \lesssim 1\,.
\eeq
On the other hand, the dimensionality of topological fluctuations is still not firmly established
and is plagued by theoretical uncertainties. Various estimations made both in this paper and in the
literature~\cite{Horvath:2005rv,Aubin:2004mp,Ilgenfritz:2005hh,Horvath:2005cv,Gubarev:2005jm,Zakharov:2006vt}
suggest that it is of order three
\beq
\label{check:D}
\mathrm{dim}[V_\pm] ~\approx ~  3\,.
\eeq
It is remarkable that Eqs.~(\ref{check:q}), (\ref{check:D}) overlap only at one point
consistent with Eq.~(\ref{dimension:relation})
\beq
\label{check:final}
\alpha ~=~ 1\,, \qquad \mathrm{dim}[V_\pm] ~=~  3\,.
\eeq
Given that Eqs.~(\ref{check:main1}), (\ref{check:main2}) are fulfilled with amazingly high accuracy
we are forced to the conclusion that (\ref{check:final}) is the only values consistent with both
theoretical considerations and numerical data.

\section{Conclusions}
\label{section:conclusions}
In this paper we further developed the $SU(2)$ gauge fields topology investigation method,
based on the embedding of $\HP{1}$ $\sigma$-model into the given gauge
background~\cite{Gubarev:2005rs,Boyko:2005rb}. Our prime purpose was to exploit
the remarkable properties of $\HP{1}$ projected fields found in~\cite{Boyko:2005rb},
namely, the factual absence of leading perturbative divergences and simultaneous existence of
non-trivial quadratic power correction to the gluon condensate.
It is clear that these striking features of the projected fields are to be encoded into the local structure
of the topology defining map $S^4 \to \HP{1}$ and hence should be reflected in the corresponding topological density.
The extended analysis of leading power corrections performed in this paper allows to conclude that
the topological density is likely to be linearly divergent in the continuum limit.
Note that this divergence has nothing to do
with short distance perturbative singularities and is much weaker. The divergence of the topological density
by itself is almost academical problem since it is not directly observable. However, combined
with the requirement of ultraviolet finiteness of topological susceptibility it
leads to rather dramatic consequences for the geometry of relevant topological fluctuations.
Namely, we argued that the topological
charge is to be concentrated in three-dimensional submanifolds of four-dimensional
Euclidean space. This is the only conclusion compatible with physical topological susceptibility,
numerically established pattern of power corrections and which does not 
necessitates explicit fine tuning of the Yang-Mills theory at UV scale. 
Moreover, the fine tuning assumption is testable and if it does not happen,
then one could derive rather stringent relation between the divergence
of the topological density and the dimensionality of submanifolds, which support the most of the
topological charge. Note, however, that the lower dimensionality of topological fluctuations
could also be considered as a sort of fine tuning, in which the explicit powers of UV cutoff
are traded for unusual geometric properties. Qualitatively our results are in accord with modern trends
in the literature, which discuss the lower dimensionality of physically relevant vacuum
fluctuations~\cite{Kovalenko:2005zx,Gubarev:2002ek}
and, in particular, of the topological density sign-coherent
regions~\cite{Horvath:2005rv,Aubin:2004mp,Ilgenfritz:2005hh,Horvath:2005cv,Gubarev:2005jm,Zakharov:2006vt,Kovalenko:2005rz}.

The actual experimental verification of this scenario turned out to be rather intricate
both conceptually and technically; we believe that all these problems were adequately
addressed in our paper. The technical achievement is the development of fast and rather precise
numerical algorithm of topological density evaluation, which allowed us to investigate the problem
on the convincing statistical level. While the UV behavior of the topological density
could be studied directly, the dimensionality of relevant topological fluctuations is much more
involved problem, which consists essentially in physical interpretation of the term
``relevant'' above. We argued that the natural approach is to embed a dynamical system
into the topological density background, the evolution of which is sensitive to the dimensionality
of ambient space. The simplest system of this sort could be constructed on the top of usual random walk model
and depends upon one dimensionless parameter. Then the phase structure of the system and
location of its critical-like points is ultimately related to the long-range properties
of the underlying background and, in particular, to the dimensionality of sign-coherent topological regions.

As far as the results of numerical experiments are concerned, our data show unambiguously that non-perturbatively
defined characteristic topological density is divergent in the continuum limit.
Moreover, we were able to obtain the upper bound on its leading spacing dependence.
At the same time, the dimensionality $D$ of the relevant
topological fluctuations was shown to be decidedly less than four 
and is compatible with $D=3$.
Here the assumed absence of the fine tuning becomes crucial and we showed that it indeed does not happen.
Instead the topological charges associated with sign-coherent regions fluctuate independently.
We conclude therefore that the only possibility to satisfy all the restrictions is to have linearly
divergent topological density distributed in three-dimensional domains.

\section*{Acknowledgments}
The authors are grateful to Prof.~V.I.Zakharov and to the members of ITEP lattice group
for stimulating discussions. The work was partially
supported  by grants RFBR-05-02-16306a and RFBR-05-02-17642.
F.V.G. was partially supported by INTAS YS grant 04-83-3943.

\renewcommand{\theequation}{A.\arabic{equation}}
\setcounter{equation}{0}
\section*{Appendix}
Here we describe in details the numerical algorithm used in this paper to calculate the topological charge density.
The general formulation of the problem could be found in Refs.~\cite{Gubarev:2005rs,Boyko:2005rb}.
Essentially it reduces to the evaluation of the volume $V(T)$ of 4-dimensional spherical tetrahedron $T$
embedded into $S^4$ with vertices $n^A_i$ given in terms of five $i = 0, ..., 4$ unit five-dimensional
$A = 0, ..., 4$ vectors, $(\vec n_i)^2 = 1$.

It is clear that for near vanishing $V(T)$ the volume is given by $V(T) = \mathrm{det}_{Ai}[n^A_i]$.
Thus the volume of finite tetrahedron could be found by triangulating it into the set of small tetrahedra
and summing up the infinitesimal volumes.
In fact, our method works recursively until the determinant estimation of the volume
of input tetrahedron is larger than $10^{-6}$; this way we indeed obtain the optimal performance.
However, the determinant-based volume estimation is reliable and uniform only if the angles between all input
vertices are small enough, e.g. $(\vec n_i, \vec n_j) > 0$. In this case it suffices to place
new triangulation vertex at $\vec m \propto \sum_i \vec n_i$ and complete the recursion cycle.
However, this procedure must be modified if there are at least two input vertices $i$ and $j$
for which $(\vec n_i, \vec n_j) < 0$.
Indeed, in this case the above recursions converge non-uniformly and eventually lead to almost
degenerate tetrahedra with large volume but still almost vanishing determinant.
The needed modification is to take the new vertex at $\vec m \propto \vec n_i + \vec n_j$ which guarantees
that eventually we will get $(\vec n_i, \vec n_j) > 0$ $\forall i,j$.

\begin{figure}[t]
\centerline{\psfig{file=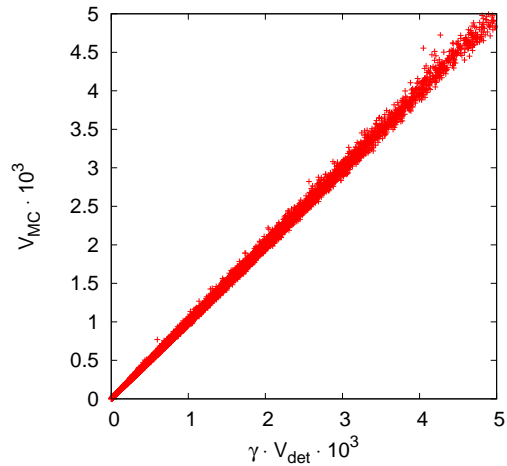,width=0.5\textwidth,silent=,angle=-90}}
\caption{Cumulative distribution of $V_{MC}(T)$ and $V_{det}(T)$ (see text).}
\label{fig:appendix}
\end{figure}

Let us note that for given accuracy of the determinant volume estimation for small tetrahedra
(which is $10^{-6}$ in our case) the volume of the original tetrahedron is evaluated, in fact, with
finite bias which is due to the sphericity of every small tetrahedron. However for small enough
volumes the sphericity could be accounted for by simple rescaling of the determinant estimation.
To calibrate the present algorithm we compared it with our previous Monte-Carlo based 
method. To this end we generated $5 \cdot 10^4$ random spherical tetrahedra and applied both
algorithms to each of them thus obtaining Monte-Carlo $V_{MC}(T)$ and determinant-based $V_{det}(T)$
volume estimations. Fig.~\ref{fig:appendix} represents  the cumulative distribution of
$V_{MC}(T)$, $V_{det}(T)$, which turns out to be astonishingly narrow and is fairly compatible
with linear dependence
\beq
\label{appendix:fit}
V_{MC}(T) ~=~ \gamma \cdot V_{det}(T)\,, \qquad \gamma ~=~ 1.1285(5)\,,
\eeq
where the optimal value of $\gamma$ coefficient results from the best linear fit. 
Note that the plot on Fig.~\ref{fig:appendix}  is restricted to $V(T) \lesssim 5 \cdot 10^{-3}$,
which is far beyond the maximal value of topological density even for our largest spacing;
however, the linear dependence (\ref{appendix:fit}) remains valid even at larger $V(T)$.
Thus we are confident that the new triangulation method is definitely
compatible with old Monte-Carlo approach  in the relevant range of lattice spacings.

Finally, we performed the same check as one described in \cite{Boyko:2005rb}.
Namely, we confronted the global
topological charge, which could be found unambiguously for each our configuration
with one calculated with present algorithm. It turns out
that they agree in all cases with no exceptions. Moreover, we found that the determinant-based algorithm
gives even narrower distribution of $Q_{float}$ around integer numbers compared to that of old Monte-Carlo based
approach. We conclude therefore that the new method of the topological density calculation
is superior to the old one both in accuracy and performance.


\end{document}